\documentclass[twocolumn,aps,showpacs,prx,superscriptaddress,preprintnumbers,amsmath,amssymb]{revtex4-2}
\usepackage{graphicx}% Include figure files
\usepackage{dcolumn}% Align table columns on decimal point
\usepackage{appendix}
\usepackage{bm}% bold math

\usepackage[usenames,dvipsnames]{xcolor}

\usepackage{subfigure}
\usepackage{bbm}
\usepackage{enumerate}

\usepackage[
  bookmarks=true,
  colorlinks,
  linkcolor=black,
  urlcolor=black,
  citecolor=black,
  plainpages=false,
  pdfpagelabels,
  final,
  breaklinks=true
]{hyperref}

\usepackage{titlesec}

\usepackage{array}
\usepackage{dsfont}

\usepackage{physics}

\begin{document}

\title{Theory of entanglement and measurement in high harmonic generation}

\author{Philipp Stammer}
\email{philipp.stammer@icfo.eu}
\affiliation{ICFO -- Institut de Ciencies Fotoniques, The Barcelona Institute of Science and Technology, 08860 Castelldefels (Barcelona), Spain}

\date{\today}

\begin{abstract}

Quantum information science and intense laser matter interaction are two apparently unrelated fields. 
%However, the recent developments of the quantum optical description of the intense laser driven process of high harmonic generation allow to conceive new light engineering protocols. 
Here, we introduce the notion of quantum information theory to intense laser driven processes by providing the quantum mechanical description of measurement protocols for high harmonic generation in atoms. 
This allows to conceive new protocols for quantum state engineering of light.
We explicitly evaluate conditioning experiments on individual optical field modes, and provide the corresponding quantum operation for coherent states. 
The associated positive operator-valued measures are obtained, and give rise to the quantum theory of measurement for the generation of high dimensional entangled states, and coherent state superposition with controllable non-classical features on the attosecond timescale.
This establish the use of intense laser driven processes as a novel quantum technology platform for quantum information processing.

\end{abstract}

\maketitle
%\tableofcontents
%\begin{multicols}{2}

%\section{\label{sec:intro}Introduction}

\textit{Introduction.}
In the process of high harmonic generation (HHG), coherent radiation of higher order harmonics of the driving laser frequency is generated \cite{lewenstein1994theory, salieres1997study}. 
The transfer of coherence, and energy, from the intense laser source to the harmonic field modes (initially in the vacuum) is achieved by a highly nonlinear interaction of the driving field with the HHG medium, in which the electron is used as an intermediary between the optical modes.
Until recently this was mainly described in semi-classical terms, in which only the electronic degrees of freedom are quantized \cite{lewenstein1994theory}, although there have been early approaches to introduce a fully quantized description of the HHG process \cite{sundaram1990high, eberly1992spectrum, xu1993non, compagno1994qed, gauthey1995role, becker1997unified, diestler2008harmonic}. However, recent advances in the quantum optical analysis of HHG  has established a new direction in the investigation of strong field physics. This allows to study the quantum mechanical properties of the harmonic radiation, or to take into account the backaction on the driving field \cite{kominis2014quantum, gonoskov2016quantum, tsatrafyllis2017high, tsatrafyllis2019quantum, gombkotHo2020high, gorlach2020quantum, lewenstein2021generation, rivera2021quantum, stammer2021high, rivera2021new, stammer2022quantum, lewenstein2022attosecond, rivera2022light, gombkotHo2021quantum}. 
In particular, it was shown that conditioning procedures on processes induced by intense laser-matter interaction can lead to the generation of high-photon number controllable non-classical field states in a broad spectral range \cite{lewenstein2021generation, stammer2022quantum, rivera2021quantum, stammer2021high, rivera2021new}. 

However, a complete description of these measurement processes is missing thus far, and to make use of the full power of such protocols for quantum state engineering in optical quantum technologies \cite{o2009photonic, walmsley2015quantum, acin2018quantum} the corresponding theory is needed.
% and the properties of the states of light are needed.
One goal in this direction is to construct the quantum theory of measurement for HHG, and to obtain the generic structure of the entangled states of the electron and the field modes in strong field processes.
Both is presented in the present work by introducing the natural language in which measurements are described.
This is done by providing the necessary measurement operations, and corresponding positive operator-valued measure \cite{kraus1983states, nielsen2002quantum}.
The notion of measurement operations is well known in quantum information science, and has lead to many advances ranging from light-engineering protocols \cite{gerry2005introductory}, describing open quantum systems \cite{breuer2002theory} or quantum control \cite{wiseman2009quantum}. But, these methods play a new role in the description of the interaction of an intense laser field with matter.
A quantum operation $\Phi_n(\cdot)$ is a completely positive map of one quantum state to another $\rho \mapsto \rho_n = \Phi_n(\rho)/P(n)$, and describe how the state $\rho$ changes during a process with outcome $n$, which occurs with probability $P(n) = \Tr[\Phi_n(\rho)]$. The representation theorem for quantum operations states that the operation is expressed as $\Phi_n(\rho) = M_n \rho M_n^\dagger$, with linear operators $M_n$ satisfying $\sum_n M_n^\dagger M_n = \mathds{1}$ \cite{kraus1983states}. The operators $M_n$ correspond to generalized measurement operations, which for the special case of projective (von Neumann) measurements are given by projection operators $M_n = \dyad{n}$. This concept of quantum operations naturally leads to the notion of measurement theory via positive operator-valued measure (POVM), in which each possible measurement outcome $n$ is associated to a positive operator $E_n = M_n^\dagger M_n$. These operators are the elements of the set $\{ E_n \}$ defining the POVM, and allow to construct all possible operations which can be performed on a quantum system. 
It will be shown, that within this framework, conditioning measurements on the electron allow to generate massive, and controllable, entangled and superposition states of the optical field in intense laser matter interaction. 

The main goal of the present work is to ultimately connect the domain of strong field physics with quantum information science by introducing the relevant quantum formalism of measurements. 
This allows to derive the quantum operations for the measurement scheme introduced in \cite{stammer2021high}, and to proof that the corresponding measurement operators are elements of a POVM. Is is further shown under which conditions these generalized measurement operations reduce to the projection operators phenomenologically introduced in \cite{lewenstein2021generation, rivera2021quantum} to describe the generation of optical cat states.
The derived measurement operations for conditioning schemes in HHG allow to perform quantum state engineering protocols, to anticipate the generation of novel stats of light with previously inaccessible photon numbers, frequencies, and timescales.

%We provide the theory of measurement for conditioning schemes in HHG to perform quantum state engineering protocols, which allows to generate novel states of light with previously inaccessible photon numbers, frequencies, and timescales. 

%With this we can describe the recent experimental observation of an optical cat-state \cite{lewenstein2021generation} by means of the POVM formalism, and allows access the underlying structure of the entangled field state. 

%The main findings of the present work are: (i) We provide the detailed description of the entangled state of the optical field modes in HHG with the necessary quantum operation to separate it from idle processes. (ii) We introduce a complete quantum theory of measurement for conditioning schemes in HHG to perform quantum state engineering protocols. (iii) This ultimately connects the domain of strong field physics with quantum information science, which allows to generate novel states of light with previously inaccessible photon numbers, frequencies, and timescales.

%\section{Entanglement in strong field physics}
\textit{Entanglement in strong field physics.}
To describe the entangled state of the total system including the electron of a single atom and the optical field after strong field light matter interaction we consider an initially uncorrelated state $\ket{\psi(t_0)} = \ket{\phi_i} \ket{g}$. The initial state of the optical field $\ket{\phi_i} = \otimes_q \ket{\phi_q}$ is assumed to be pure, and that the modes $q$ are uncorrelated, while the electron is initially in the ground state $\ket{g}$. 
The formal solution of the Schr{\"o}dinger equation describing the interaction is given by
\begin{align}
\label{eq:solution_TDSE}
    \ket{\psi(t)} = U(t,t_0) \ket{\phi_i} \ket{g}.
\end{align}

In general, the field and the atom are correlated after the interaction, and neither the total state $\ket{\psi(t)} \neq \ket{\phi(t)} \otimes \ket{\epsilon(t)}$, where $\ket{\epsilon(t)}$ describes the state of the electron, nor the optical field modes $\ket{\phi(t)} \neq \otimes_q \ket{\phi_q(t)}$ are described as a product state after the interaction.
To make the correlation between the field and the electron more apparent we introduce the identity on the electronic subspace 
\begin{align}
    \mathds{1} = \dyad{g} + \sum_\mu \dyad{\lambda_\mu} +  \int d\vb{v} \dyad{\vb{v}},
\end{align}
where $\ket{\lambda_\mu}$ are the excited electronic bound states, and $\ket{\vb{v}}$ are the continuum states with electron momentum $\vb{v}$. The state \eqref{eq:solution_TDSE} can now be written as 
\begin{align}
\label{eq:entanglement_field_atom1}
    \ket{\psi(t)} = & K_{HHG} \ket{\phi_i} \ket{g} + \sum_\mu K_{EX}^\mu  \ket{\phi_i} \ket{\lambda_\mu}  \\
    & + \int d\vb{v} K_{ATI}(\vb{v}) \ket{\phi_i} \ket{\vb{v}}, \nonumber
\end{align}
where $K_{HHG} = \bra{g} U(t,t_0) \ket{g}$, $K_{EX}^\mu = \bra{\lambda_\mu} U(t,t_0) \ket{g}$, and $K_{ATI}(\vb{v}) = \bra{\vb{v}} U(t,t_0) \ket{g}$ are the operators solely acting on the optical field state. 
The first term corresponds to the electron conditioned on the ground state $\ket{g}$ leading to HHG \cite{lewenstein2021generation}, while the third term corresponds to an electron in a continuum state $\ket{\vb{v}}$ leading to the process of above-threshold ionization (ATI) \cite{becker2002above, rivera2022light}. The second term, corresponding to the electron being in an excited bound state $\ket{\lambda_\mu}$ is usually considered to be small compared to the contribution of HHG and ATI. However, bound state resonances can lead to important contributions in strong field ionization \cite{freeman1987above, perry1989resonantly, stammer2020evidence} or high harmonic generation \cite{jimenez2017time, mayer2021role}, and are therefore included in the present analysis.  
Hence, the interaction of the intense laser field with the atom leads to the entangled state \eqref{eq:entanglement_field_atom1} in which the electron is either found in the ground state, in an excited state or in a continuum state after the interaction
\begin{align}
    \ket{\psi} = \ket{\phi_g} \ket{g} + \sum_\mu \ket{\phi_\mu} \ket{\lambda_\mu} + \int d\vb{v} \ket{\phi_{\vb{v}}} \ket{\vb{v}},
\end{align}
where $\ket{\phi_g} = K_{HHG} \ket{\phi_i}$, and $\ket{\phi_{\vb{v}}} = K_{ATI}(\vb{v}) \ket{\phi_i}$ are the optical field states corresponding to the electron conditioned on HHG or ATI, respectively.
The term corresponding to an electronic excitation is associated to the field in the state $\ket{\phi_\mu} = K_{EX}^\mu \ket{\phi_i}$.
This is the generic entangled state between the optical field and the electron in intense laser matter interaction. We only assumed an initial pure state of the field in tensor product form, but no further specification was introduced. The state is written in the form where experimental conditioning schemes on the matter part (electronic ground state, excitation or ionization) becomes apparent. 

\textit{Entanglement in high harmonic generation.}
Since HHG is one of the most prominent processes in intense laser matter interaction, we shall now describe the entanglement between the different optical field modes, i.e. between the fundamental and the harmonic modes, in more detail. 
To account for the process of HHG a conditioning measurement on the electron has to be performed. Namely, we project onto the electronic ground state $\Pi_{HHG} = \dyad{g}$. This can, for instance, be done by considering only the processes where no electron was detected, and assuming that the excited state population is negligible.
In all remaining events, the electron is in the ground state, and the corresponding state is given by $\Pi_{HHG} \ket{\psi} = \ket{\phi_g} \ket{g}.$
This conditioning effectively eliminates all idle processes which do not correspond to HHG. 
Since the object of interest is the optical field state $\ket{\phi_g}$, we trace over the electronic degrees of freedom. 
To describe the entanglement between the different optical field modes we shall now analyze the optical state in more detail 
\begin{align}
\label{eq:field_HHG_uncorr}
    \ket{\phi_g} = K_{HHG}\ket{\phi_i} = \bra{g} U(t,t_0) \ket{g} [\otimes_q \ket{\phi_q} ].
\end{align}

To model the HHG process we need to solve $K_{HHG}$, which is done by assuming that the intense driving laser can be described by a single mode pure coherent state $\ket{\phi_1} = \ket{\alpha}$, while the harmonic modes $q \in \{ 2, ..., N \}$ are initially in the vacuum $\ket{\{ 0_q \}} = \otimes_{q \ge2}\ket{0_q}$ such that the initial field state is given by $\ket{\phi_i} = \ket{\alpha} \otimes \ket{\{0_q \}}$. The assumption of having a pure coherent state for the driving field mode implies that experimental realizations need to operate with a phase-stabilized laser source. A single mode field to describe the pulsed laser is used since the measurement scheme does not discriminate between the individual modes of the driving field. 
The electron is assumed to be initially in the ground state $\ket{g}$. The Hamiltonian describing the interaction of a single electron with the optical field in the length-gauge, and within the dipole approximation is given by $H_I(t) = - \vb{d}(t) \cdot \vb{E}$, where the electric field operator $\vb{E}(t) = - i \vb{\kappa} \sum_{q=1}^N \sqrt{q} \left( b_q^\dagger e^{i q \omega t} - b_q e^{- i q \omega t}  \right)$ is coupled to the dipole moment operator $\vb{d}(t) = U^\dagger_{sc}(t,t_0) \vb{d} U_{sc}(t,t_0)$. The time-dependent dipole moment operator is in the interaction picture of the semi-classical frame $U_{sc}(t,t_0)= \mathcal{T} \exp[- i \int_{t_0}^t d\tau H_{sc}(\tau) ]$, with respect to the Hamiltonian $H_{sc}(t) = H_a - \vb{d} \cdot \vb{E}_{cl}(t)$. This is exactly the Hamiltonian for the semi-classical description of intense laser fields interacting with a single electron, where $H_a = \vb{p}^2/2 + V(\vb{r})$ is the pure atomic Hamiltonian, and $\vb{E}_{cl}(t) = \bra{\phi_i} \vb{E}(t) \ket{\phi_i} = i \kappa (\alpha e^{- i \omega t} - \alpha^* e^{i \omega t})$ is the classical part of the initial optical field. A detailed derivation of the interaction Hamiltonian $H_I(t)$ can be found in \cite{lewenstein2021generation, rivera2021quantum}.
%The time evolution of the total system governed by the Hamiltonian $H_I(t)$ is given by $U(t,t_0) = \mathcal{T} \exp{- i \int_{t_0}^t dt^\prime H_I(t^\prime)}$. 
To now describe the time evolution of the optical field governed by the Hamiltonian $H_I(t)$, and conditioned on HHG, it remains to solve
\begin{align}
    K_{HHG} = \bra{g} \mathcal{T} \exp{ i \int_{t_0}^t dt^\prime \vb{d}(t^\prime) \cdot \vb{E}(t^\prime) } \ket{g}.
\end{align}

Assuming that there are no correlations between dipole moments \cite{sundaram1990high}, it was shown \cite{lewenstein2021generation, rivera2021quantum}, that this can be approximated by coupling the dipole moment expectation value to the electric field instead of the dipole moment operator itself, such that
%It was shown that this can be well approximated when neglecting the electric correlations of the dipole moment \cite{sundaram1990high, lewenstein2021generation, rivera2021quantum}, and hence, instead of the dipole moment operator, the dipole moment expectation value is coupled to the electric field
\begin{align}
\label{eq:kraus_approx}
    K_{HHG} \simeq \mathcal{T} \exp{i \int_{t_0}^t dt^\prime \expval{\vb{d}}(t^\prime) \cdot \vb{E}(t^\prime) },
\end{align}
where $\expval{\vb{d}}(t) = \bra{g} \vb{d}(t) \ket{g} = \bra{\epsilon(t)} \vb{d} \ket{\epsilon(t)}$ is the dipole moment expectation value. The electronic state $\ket{\epsilon(t)} = U_{sc}(t,t_0) \ket{g}$ is given by the conventional solution of HHG in the semi-classical description \cite{lewenstein1994theory}. 
The operator in \eqref{eq:kraus_approx} can now be obtained exactly by noting that the electric field operator $\vb{E}(t)$ is linear in the creation- and annihilation operators, and the solution is given by a multimode displacement operator $D(\chi ) = \exp(\chi b^\dagger - \chi^* b)$ \cite{lewenstein2021generation, stammer2021high, rivera2021quantum, stammer2022quantum}
\begin{align}
    K_{HHG} = \prod_{q=1}^N e^{i \varphi_q}D(\chi_q),
\end{align}
where the shift of the fundamental mode and the harmonic modes account for the depletion of the driving laser amplitude due to the generation of coherent radiation in the harmonic modes with amplitudes $\chi_q$. The state (up to a phase) is then given by
%The amplitude of the displacement in phase-space for each mode $q$ is given by the respective Fourier component of the oscillating time dependent dipole moment 
%\begin{align}
%\label{eq:shift}
%    \chi_q = - i\sqrt{q} \vb{\kappa} \int_{-\infty}^\infty dt^\prime \expval{\vb{d}}(t^\prime) e^{i q \omega t^\prime},
%\end{align}
%where we have considered $t_0 \to -\infty$, and the asymptotic limit $t \to \infty$.
%The phase shift of each mode is given by 
%\begin{align}
%    \varphi_q = q  \vb{\kappa}^2 \int_{-\infty}^\infty dt_1 \int_{-\infty}^{t_1} dt_2 \vb{d}(t_1)  \vb{d}(t_2) \sin[q \omega (t_1-t_2)].
%\end{align}
%To model the HHG process we assume a single mode intense driving laser, which we describe by a coherent state $\ket{\phi_1} = \ket{\alpha}$, while the harmonic modes $q \in \{ 2, ..., N \}$ are initially in the vacuum $\ket{\{ 0_q \}} = \otimes_{q \ge2}\ket{0_q}$. The field amplitude of the driving laser is typically in the range of $\abs{\alpha} \sim 10^7$.
%The interaction of the field with the atomic medium conditioned on the electronic ground state (leading to HHG) can then be described as an effective multimode displacement operation (see Methods) \cite{lewenstein2021generation, stammer2021high, rivera2021quantum}
\begin{align}
\label{eq:field_shifted}
    \ket{\phi_g} &= \prod_{q=1}^N D(\chi_q) \ket{\phi_i}  = \ket{\alpha + \chi_1} \bigotimes_{q=2}^N \ket{\chi_q},
\end{align}
and the shift of the individual field modes $\chi_q = - i  \vb{\kappa} \sqrt{q} \expval{\vb{d}}(q\omega)$ is given by the respective Fourier component of the time dependent dipole moment expectation value $\expval{\vb{d}} (q \omega) = \int dt \expval{\vb{d}}(t) e^{i q \omega t}$. 
%with the coupling constant $\vb{\kappa}$, which takes into account the field polarization, and the Fourier component of the time dependent dipole moment expectation value $\expval{\vb{d}} (q \omega) = \int_{- \infty}^\infty dt \expval{\vb{d}}(t) e^{i q \omega t}$. 
%The shift of the fundamental mode $\chi_1$ accounts for the depletion of the driving laser due to the generation of coherent radiation in the harmonic modes with amplitudes $\chi_q$.
The coherent properties of the fundamental laser are transferred to the harmonic modes by virtue of the non-linear interaction with the electron, and that the harmonic modes are described by coherent states is due to the fact that the source for the coherent radiation is related to the electron dipole moment expectation value which acts as a classical charge current.
Note that we have so far considered a single atom interacting with the intense laser field, which is applicable under the assumption of vanishing interatomic correlations \cite{sundaram1990high}. However, the concepts and methods introduced in the following are applicable for generic $\chi_q$, and extending to the more realistic scenario where many atoms contribute to the HHG process we can multiply the displacement $\chi_q$ by the number of atoms which contribute in a phase-matched way \cite{lewenstein2021generation}. 
Furthermore, macroscopic effects on the high harmonic generation mechanism itself, e.g. focal averaging, can be included in the computation of the time-dependent dipole moment expectation value $\expval{\vb{d}}(t)$. This is also where the microscopic details of the HHG medium is included, and sophisticated numerical or analytical tools known from semi-classical HHG theory are needed \cite{armstrong2021dialogue}.

However, the induced time-dependent dipole moment $\expval{\vb{d}(t)} = \bra{g} \vb{d}(t) \ket{g}$ makes the respective shifts of the modes correlated. In fact, the actual mode which is populated during the HHG process is a wavepacket mode taking into account these correlations \cite{lewenstein2021generation, stammer2021high}, and can be described by the corresponding number states $\ket{\tilde n}$.
To account for the process of HHG, we express the state in \eqref{eq:field_shifted} in terms of the number states of the wavepacket mode 
\begin{align}
\label{eq:rho_basis_wavepacket}
    \ket{\phi_g} = \sum_{\tilde n} \bra{\tilde n} \ket{\phi_g} \ket{\tilde n}.
\end{align}

We can now consider two mutual exclusive events, which is either emitting harmonic radiation or not. If harmonic radiation is generated the wavepacket described by $\ket{\tilde n}$ is excited, and of course, if no harmonics are generated the wavepacket is in its vacuum state $\ket{\tilde 0}$. We can therefore define the set of operators 
\begin{align}
\label{eq:POVM_wavepacket}
    \mathcal{A}_{\tilde n} = \{ \Pi_{\tilde 0}, \, \Pi_{\tilde n \neq 0} \}, 
\end{align}
which either projects on the sub-space of all possible excitation of the wavepacket mode
\begin{align}
\label{eq:def_1-minus-initial}
\Pi_{\tilde n \neq 0} = \sum_{\tilde n \neq 0} \dyad{\tilde n},     
\end{align}
or on its vacuum state
\begin{align}
    \Pi_{\tilde 0} = \dyad{\tilde 0}.
\end{align}

This constitute the POVM elements for conditioning between the generation of harmonic radiation, and no excitation of the corresponding wavepacket mode. 
We shall now consider the more interesting case in which harmonic radiation is generated, i.e. where an excitation of the wavepacket mode is present. Projecting on $\Pi_{\tilde n \neq 0}$ we obtain 
\begin{align}
\label{eq:rho_wavepacket}
    \ket{\phi_{HH}} = \Pi_{\tilde n \neq 0} \ket{\phi_g} =  \sum_{\substack{\tilde n \neq 0}} \bra{\tilde n} \ket{\phi_g} \ket{\tilde n}.
\end{align}

We use that $ \sum_{\tilde n \neq 0} \dyad{\tilde n} = \mathds{1} - \dyad{\tilde 0}$, and that the vacuum state of the wavepacket mode is given by the state of the field prior to the interaction with the atomic  medium 
\begin{align}
    \ket{\tilde 0} = \ket{\alpha} \bigotimes_{q=2}^N \ket{0_q},
\end{align}
such that the state in \eqref{eq:rho_wavepacket} is given by
\begin{align}
\label{eq:HHG_entangled_state}
    \ket{\phi_{HH}} = \ket{\alpha + \chi_1} \bigotimes_{q=2}^N \ket{\chi_q} - \xi \ket{\alpha} \bigotimes_{q=2}^N \xi_q \ket{0_q},
\end{align}
where $\xi = \bra{\alpha} \ket{\alpha + \chi_1}$, and $\xi_q = \bra{0_q} \ket{\chi_q}$. 
This is the entangled state between all modes of the optical field in the process of HHG, and is heralded by the emission of harmonic radiation. The number of modes which are entangled is on the order of the harmonic cutoff $\mathcal{O}(N)$, and presents a massively entangled state of many field modes ($N>10$). A first characterization of this state in terms of the linear entropy for different partitions can be found in \cite{stammer2021high}.
Note that this entangled state is naturally generated by the emission of harmonic radiation, and no particular measurement has to be performed on the optical modes if ionization and excitation events are eliminated. The generation of harmonic radiation by itself serves as the origin of this entangled optical field state due to the interaction with the atomic medium, and the corresponding correlations induced between the different field modes.
However, the crucial part is to experimentally separate the process of HHG from all other idle processes such as ionization or excitation, i.e. implementing the projection $\Pi_{HHG}$. This can be done by measuring the electrons which might get ionized, and neglecting all cases in which an electron was detected, i.e. only considering the cases where no ionization takes place. Alternatively, as done in \cite{lewenstein2021generation, rivera2021quantum}, the harmonic radiation can be measured together with a subsequent post-processing procedure in which only those events are taken into account in which the shift of the harmonic intensity is anti-correlated to the intensity of the fundamental mode. This likewise allows to separate HHG from all idle processes. The drawback of this scheme is that some fraction of the optical field needs to be detected, and is thus inevitably interacting with additional optical devices leading to decoherence and amplitude reduction. Nonetheless, this scheme is very powerful to generate non-classical optical cat states with high photon numbers \cite{lewenstein2021generation, stammer2021high, rivera2021quantum, rivera2021new}.

%\section{Quantum state engineering of light using HHG}

%\subsection{\label{sec:measurement_QHHG}Quantum operations in HHG}
\textit{Quantum operations in HHG.}
In the following we introduce the quantum theory of measurement to describe conditioning experiments in HHG. This can be seen as a new class of quantum state engineering protocols of light. 
Suppose we perform a post-selection measurement on the state $\rho_{HH} = \dyad{\phi_{HH}}$ from HHG \eqref{eq:HHG_entangled_state}. In particular, we perform a measurement on $M$ field modes (denoted by $q^\prime$), and post-select on the coherent states $\ket{\{ \tilde \chi_{q^\prime} \}}$. The state of the remaining $N-M$ field modes accordingly reads $\rho^\prime = \bra{ \{ \tilde \chi_{q^\prime} \}} \rho_{HH} \ket{\{ \tilde \chi_{q'} \}}$. 
Let us therefor define a quantum operation $\Phi_{\tilde n \neq 0}^{\tilde \chi}[\cdot]$, which corresponds to this post-selection measurement on the modes $q^\prime$, and is acting on the initial state of the remaining modes described by $\rho_0 =\Tr_{q^\prime}[\dyad{\phi_i}]$. This quantum operation for conditioning in the process of HHG reads
\begin{align}
\label{eq:operation_general}
    \rho_0 \mapsto \rho^\prime = \frac{\Phi_{\tilde n \neq 0}^{\tilde \chi}\left[ \rho_0\right]}{P_{\tilde n \neq 0}^{\tilde \chi}} = \frac{M_{\tilde n \neq 0}^{\tilde \chi} \rho_0 \left( M_{\tilde n \neq 0}^{\tilde \chi} \right)^\dagger}{\Tr\left[ E_{\tilde n \neq 0}^{\tilde \chi} \rho_0 \right]},
\end{align}
where we have defined the effect
\begin{align}
     E_{\tilde n \neq 0}^{\tilde \chi} = \left( M_{\tilde n \neq 0}^{\tilde \chi} \right)^\dagger M_{\tilde n \neq 0}^{\tilde \chi},
\end{align}
with measurement operator 
\begin{align}
    M_{\tilde n \neq 0}^{\tilde \chi} = \bra{\{ \tilde \chi_{q'} \}} \Pi_{\tilde n \neq 0} \ket{\{ \chi_{q'} \}} \prod_{q\neq q^\prime} D(\chi_q).
\end{align}

\begin{figure}
    \centering
	\includegraphics[width = 1\columnwidth]{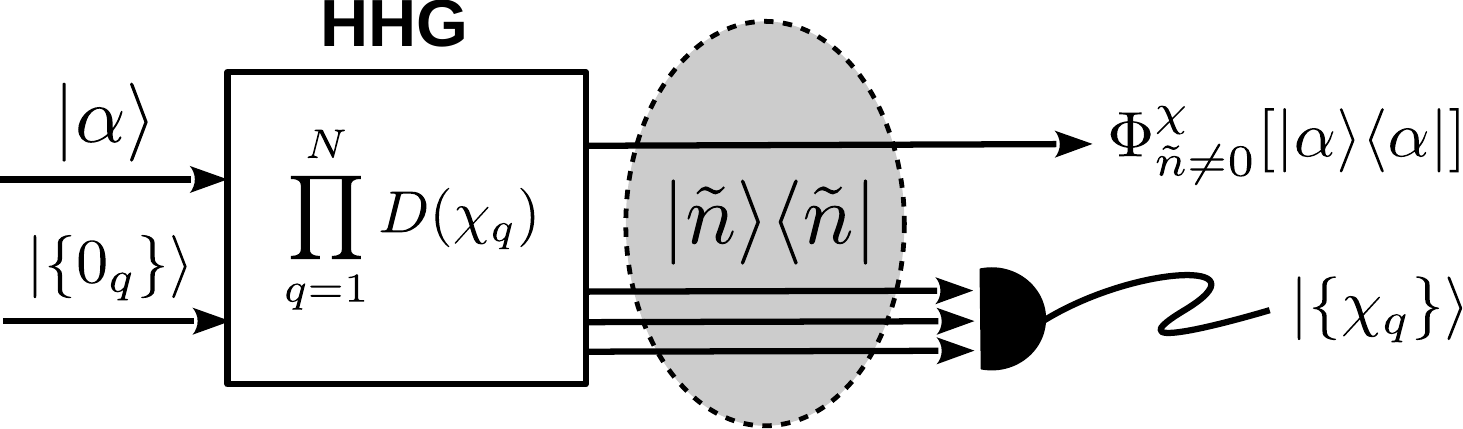}
	\caption{Schematic illustration of the conditioning experiment in HHG. The initial product state $\ket{\alpha} \otimes \ket{ \{ 0_{q\ge 2}\} }$ of the optical field mode is shifted via $K_{HHG} \simeq \prod_q D(\chi_q)$, and correlated (via the wavepacket mode $\ket{\tilde n}$).
	Performing a conditioning measurement on the harmonic modes $\ket{\{ \chi_q \}}$ leads to the quantum operation $\Phi_{\tilde n \neq0}^\chi[\cdot] $ on the initial coherent state of the driving laser.}
      \label{fig:experiment}
\end{figure}

This measurement operator takes into account the shift of the field modes due to the interaction with the HHG medium via $D(\chi_q)$, and the conditioning in terms of $\Pi_{\tilde n \neq 0}$ from the POVM in \eqref{eq:POVM_wavepacket}. The schematic illustration of the experimental conditioning scheme can be seen in Fig. \ref{fig:experiment}.
This allows to define the new POVM corresponding to the possible measurement operation in terms of coherent states $\ket{\{ \tilde \chi_{q^\prime} \}}$ performed on $M$ field modes in HHG
\begin{align}
    \mathcal{X}_{HHG} = \left\{ \left. E_\nu^{\tilde \chi} = \left( M_{\nu}^{\tilde \chi} \right)^\dagger M_{\nu}^{\tilde \chi}   \right| \nu \in \mathcal{A}_{\tilde n} \right\},
\end{align}
with POVM elements $E_\nu^{\tilde \chi}$, which satisfy the completeness relation 
\begin{align}
    \frac{1}{\pi^{N-M}} \int \prod_{q'} d^2 \tilde \chi_{q'} \sum_{\nu \in \mathcal{A}_{\tilde n}} \left( M_\nu^{\tilde \chi} \right)^\dagger M_\nu^{\tilde \chi} = \mathds{1}, 
\end{align}
where we sum over $\nu \in \mathcal{A}_{\tilde n}$, the elements of the HHG wavepacket POVM in \eqref{eq:POVM_wavepacket}.
The measurement operation introduced in \eqref{eq:operation_general} is of general kind, and permits to analyze various conditioning schemes in the process of HHG. It allows to consider different initial states of the optical field, e.g. HHG driven by multiple laser modes, non-classical or mixed states. Note that for different driving laser fields the operation \eqref{eq:field_shifted} on the modes $K_{HHG} \simeq \prod_q D(\chi_q)$, due to the interaction with the HHG medium, is in general different. 
However, the quantum operation \eqref{eq:operation_general} establishes the connection of the photonic platform of HHG towards quantum information processing for light engineering protocols in terms of coherent states. In the following we will show how this can be used to generate new quantum states of light.

%\subsection{\label{sec:cluster_HHG}High dimensional entangled states}
\textit{High dimensional entangled states.}
With the general quantum operation introduced above we shall now provide the explicit scheme to generate high dimensional and controllable entangled states from HHG. 
For instance, the measurement in \eqref{eq:operation_general} can be performed on the fundamental mode only ($q^\prime = 1$), such that $\rho_\Omega = \bra{\alpha + \chi_1}\rho_{HH} \ket{\alpha + \chi_1}$. The corresponding quantum operation $\Phi_{\tilde n \neq 0}^{\alpha + \chi_1}[\cdot]$ on the harmonic modes read (up to normalization)
\begin{align}
\label{eq:operation_on_harmonics}
    \dyad{ \{0_q \} } \mapsto \rho_\Omega = \Phi_{\tilde n \neq 0}^{\alpha + \chi_1}\left[\dyad{ \{0_q \} }\right]  
    %M_{\tilde n \neq 0}^{\alpha + \chi_1} \dyad{ \{ 0_q \}} \left( M_{\tilde n \neq 0}^{\alpha + \chi_1} \right)^\dagger,
\end{align}
with the corresponding measurement operator acting on the initial vacuum state of the harmonic modes
\begin{align}
    M_{\tilde n \neq 0}^{\alpha+ \chi_1} &= \sum_{\tilde n \neq 0} \bra{\alpha + \chi_1} \ket{\tilde n} \bra{\tilde n} \ket{\alpha + \chi_1} \prod_{q = 2}^N D(\chi_q) \nonumber \\
    & = \left[ \mathds{1} - e^{-\abs{\chi_1}^2} \dyad{\{ 0_q \}} \right] \prod_{q=2}^N D(\chi_q).
\end{align}

The resulting state of the harmonic modes under this measurement operation can be expressed as the pure state $\ket{\Psi_\Omega} = M_{\tilde n \neq 0}^{\alpha + \chi_1} \ket{\{ 0_q \}}$, and is given by 
\begin{align}
\label{eq:harmonic_cluster}
\ket{\Psi_\Omega} =  \bigotimes_{q=2}^N \ket{ \chi_q } - e^{- \abs{\chi_1}^2} \bigotimes_{q=2}^N e^{- \frac{1}{2} \abs{\chi_q}^2} \ket{ 0_q}.
\end{align} 
%with $N_\Omega = [1+e^{- \Omega} (e^{-2\abs{\chi_1}^2} - 2 e^{- \abs{\chi_1}^2})]^{-1/2}$.

This is a high dimensional entangled state of the harmonic field modes. 
It can be particularly interesting when considering the regime of small harmonic conversion efficiency, i.e. $\abs{\chi_q} < 1$, such that the coherent state of a single harmonic mode can be approximated as $\ket{\chi_q} \simeq \ket{0_q} + \chi_q \ket{1_q}$. 
The state \eqref{eq:harmonic_cluster} can then be written up to $\mathcal{O}(\abs{\chi_q}^2)$, and reads
\begin{align}
    \ket{\Psi_\Omega} = \sum_{q=2}^N \chi_q \ket{0}^{\otimes(q-2)}\ket{1_q}\ket{0}^{\otimes(N-q)}.
\end{align}

Now, considering three harmonic modes $\{q,r,s\}$, the field has the form of a $W$-state \cite{dur2000three}
\begin{align}
    \ket{\Psi_{\Omega_3}} = \chi_q \ket{1_q 0_r 0_s} + \chi_r \ket{0_q 1_r 0_s} + \chi_s \ket{0_q 0_r 1_s}.
\end{align}

Since the optical field in the process of HHG can be tailored in an accurate way it allows for very precise coherent control of the light engineering protocols, and accordingly also on the multipartite entangled states.

%\subsection{\label{sec:cats_HHG}Generation of optical cat states}
\textit{Generation of optical cat states.}
To further explore the use of the quantum operation introduced above for generating non-classical field states, we consider the case in which all harmonic modes $q^\prime \ge 2$ are measured in \eqref{eq:operation_general}. The resulting operation on the initial state of the fundamental mode reads 
\begin{align}
\label{eq:measurement_map}
    \dyad{\alpha} \mapsto \rho = \Phi_{\tilde n \neq 0}^{\chi} \left[ \dyad{\alpha} \right],
\end{align}
with the measurement operator
\begin{align}
\label{eq:measurement_operator}
    M_{\tilde n \neq 0}^{\chi} & = \bra{ \{  \chi_q \} } \Pi_{\tilde n \neq 0} \ket{ \{ \chi_q \}} D(\chi_1) \nonumber \\
        &= \left[ \mathds{1} - e^{- \Omega} \dyad{\alpha} \right] D(\chi_1),
\end{align}
where $\Omega = \sum_{q=2}^N \abs{\chi_q}^2$. 
Applying this operation on the initial state of the driving laser $\Phi_{\tilde n \neq 0}^{\chi}[ \dyad{\alpha}] =  \dyad{\Psi}$, we obtain the pure state
\begin{align}
    \ket{\Psi} = \ket{\alpha + \chi_1} - \bra{\alpha} \ket{\alpha + \chi_1} e^{- \Omega} \ket{\alpha},
\end{align}
which can interpolate between an optical kitten or cat state with non-classical signatures in the corresponding Wigner function \cite{lewenstein2021generation, rivera2021quantum}. Note that similar operation can be performed on other field modes in order to generate optical cat states towards the XUV regime \cite{stammer2021high}, which shows the applicability of these conditioning experiments on HHG for the generation of non-classical field states over a broad spectral region.
The influence of the harmonic modes, which can be considered as an environment on which conditioning measurements are performed, is captured in the decoherence factor $\Omega$.
The measurement operation in \eqref{eq:measurement_operator} consists of two contributions. First, the displacement operation for the initial coherent state taking into account the depletion of the fundamental mode due to HHG, and second, the operator $P_{\tilde n}^\chi = \mathds{1} - e^{- \Omega} \dyad{\alpha}$, which reflects the conditioning processes.
This conditioning operation does not constitute a projective measurement since $(P_{\tilde n }^\chi)^2 \neq P_{\tilde n}^\chi$. But, we observe that for a small decoherence factor $\Omega$ the operator $P_{\tilde n}^\chi$ is indeed given by a projective measurement, such that $(P_{\tilde n}^\chi)^2 = P_{\tilde n}^\chi = \mathds{1}- \dyad{\alpha}$. This is the projective measurement phenomenologically introduced in \cite{lewenstein2021generation}, to interpret the measured Wigner function after an experimental conditioning procedure on HHG \cite{stammer2022quantum}.
Note that this projection is performed on the sub-space which is "orthogonal" to the initial state,  i.e. the identity subtracted by the initial state of the fundamental mode. 
In \cite{stammer2021high} it was shown that the environmental induced decoherence factor scales as $\Omega \propto \mathcal{O}(1/N)$ with the harmonic cutoff $N$. Since the harmonic cutoff usually extends towards values of $N > 10$, this scheme of generating optical cat states with large amplitudes is robust against this source of decoherence. 

However, we observe that the measurement operator \eqref{eq:measurement_operator} corresponds to the set of operators $\{ M_{\nu}^{\tilde \chi} \}$, where
\begin{align}
    M_\nu^{\tilde \chi} = \bra{ \{ \tilde \chi_q \} } \Pi_\nu \ket{ \{ \chi_q \}} D(\chi_1),    
\end{align}
which form the POVM defined by the set $\{ E_{\nu}^{\tilde \chi} \}$ with POVM elements 
\begin{align}
\label{eq:POVM}
E_{\nu}^{\tilde \chi} &= \left( M_{\nu}^{\tilde \chi} \right)^\dagger  M_{\nu}^{\tilde \chi}, 
\end{align}
satisfying the completeness relation 
\begin{align}
\frac{1}{\pi^{N-1}} \int \prod_{q=2}^N d^2 \tilde \chi_{q} \sum_{\nu \in \mathcal{A}_{\tilde n}} \left( M_{\nu}^{\tilde \chi} \right)^\dagger  M_{\nu}^{\tilde \chi} = \mathds{1}.
\end{align}
 
This proofs that the derived measurement operators $\{ M_\nu^{\tilde \chi} \}$ belong to a POVM, and provides the mathematical structure of the conditioning schemes given in \cite{lewenstein2021generation, rivera2021quantum, stammer2021high}.

%\section{\label{sec:conclusion}Discussion}
\textit{Discussion.}
We introduced the quantum theory of measurement to the process of HHG, which establishes the connection between the two mainly unrelated fields of quantum information science and intense laser matter interaction.
Using the general quantum operation in terms of coherent states allows to conceive new light engineering protocols for the generation of high dimensional, and controllable, optical entangled states. 
Using attosecond physics as a novel platform for quantum information processing brings the advantage of the intrinsic ultrafast time-scale, together with the high degree of coherent control of the HHG mechanism by complex polarization states or spatial modes of the driving field \cite{fleischer2014spin, ivanov2014high, jimenez2018control, hickstein2015non}. This allows to tailor the continuous-variable non-classical field states in the desired way for quantum information processing protocols \cite{lloyd1999quantum, jeong2001quantum, gilchrist2004schrodinger, ourjoumtsev2006generating, weedbrook2012gaussian, vlastakis2013deterministically}, quantum computation \cite{ralph2003quantum} or quantum key distribution \cite{jouguet2013experimental}.
Furthermore, the entangled coherent state can be used for quantum metrology \cite{joo2011quantum}, or in correlation experiments towards the use for fundamental tests of quantum theory by violating Bell-type inequalities \cite{gilchrist1998contradiction, gilchrist1999contradiction, wenger2003maximal, garcia2004proposal}, whereas the optical cat states allow for spectroscopy with non-classical states of light \cite{kira2011quantum, dorfman2016nonlinear, mukamel2020roadmap}.
Considering that HHG from more complex materials such as solids or plasma \cite{vampa2014theoretical, vampa2015linking, osika2017wannier, ghimire2019high, lamprou2021quantum}, strongly correlated \cite{silva2018high}, or topological systems \cite{bauer2018high, jurss2019high, silva2019topological, chacon2020circular, baldelli2022detecting} follows very similar mechanism anticipates for further investigation of the non-classical properties from the interaction with such materials.
In particular, entanglement in strong field driven processes, for instance, between the ion and ionized electron \cite{koll2022experimental}, or in two-electron ionization \cite{maxwell2021entanglement} are of current interest. 
The connection to quantum information science will
ultimately help to answer the question of the quantum mechanical properties in intense laser matter interaction from atoms to complex materials, and how new quantum states of light with the use for modern quantum technologies can be generated \cite{stammer2022quantum, lewenstein2022attosecond}.

%\begin{acknowledgments}
\textit{Acknowledgments.}
I thank Javier Rivera-Dean and Maciej Lewenstein for stimulating discussions about entanglement in strong field driven processes, and Paris Tzallas for helpful comments on the manuscript. 
This project has received funding from the European Union’s Horizon 2020
research and innovation programme under the Marie Skłodowska-Curie grant
agreement No 847517.
ICFO group acknowledges support from: ERC AdG NOQIA; Agencia Estatal de Investigación (R$\&$D project CEX2019-000910-S, funded by MCIN/ AEI/10.13039/501100011033, Plan National FIDEUA PID2019-106901GB-I00, FPI, QUANTERA MAQS PCI2019-111828-2, Proyectos de I+D+I “Retos Colaboración” QUSPIN RTC2019-007196-7);  Fundació Cellex; Fundació Mir-Puig; Generalitat de Catalunya through the CERCA program, AGAUR Grant No. 2017 SGR 134, QuantumCAT \ U16-011424, co-funded by ERDF Operational Program of Catalonia 2014-2020; EU Horizon 2020 FET-OPEN OPTOLogic (Grant No 899794); National Science Centre, Poland (Symfonia Grant No. 2016/20/W/ST4/00314); Marie Sk\l odowska-Curie grant STREDCH No 101029393; “La Caixa” Junior Leaders fellowships (ID100010434) and EU Horizon 2020 under Marie Sk\l odowska-Curie grant agreement No 847648 (LCF/BQ/PI19/11690013, LCF/BQ/PI20/11760031,  LCF/BQ/PR20/11770012, LCF/BQ/PR21/11840013).

%\end{acknowledgments}

\bibliographystyle{unsrt}
\bibliography{references}{}
%\end{multicols}

\end{document}